\documentclass[aps,pre,twocolumn]{revtex4}
\usepackage{amssymb}
\UseRawInputEncoding
\usepackage{eurosym}
\usepackage{epsfig}

\usepackage[utf8]{inputenc}

\begin{document}

\title{Notes about collapse in magnetohydrodynamics}

\author{E.A. Kuznetsov$^{a,b,c,d}$\/\thanks{%
kuznetso@itp.ac.ru} and E.A. Mikhailov$^{e}$}
\affiliation{{\small \textit{$^{a}$ P.N.Lebedev Physical Institute of RAS, Moscow, Russia\\
$^{b}$  L. D. Landau Institute for Theoretical Physics of RAS, Chernogolovka, Russia \\
$^{c}$ Skolkovo Institute of Science and Technology, Skolkovo, Russia \\
$^{d}$ Space Research Institute of RAS, Moscow, Russia\\
$^{e}$ Faculty of Physics, M.V.Lomonosov Moscow State University, Moscow, Russia} }
}

\begin{abstract}
We discuss a problem about magnetic collapse as a possible process for singularity formation of the magnetic field in a finite time within ideal magneto-hydrodynamics for incompressible  fluids. This process is very important from the point of view of various astrophysical applications, in particular, as a mechanism of magnetic filaments formation in the convective zone of the Sun. The collapse possibility is connected with  compressibility of continuously distributed magnetic field lines.
A well-known example of the formation of magnetic filaments in the kinematic dynamo approximation
with a given velocity field, first considered by Parker in 1963 \cite{parker1963kinematical}, rather indicates that the increase in the magnetic field is exponential in time. In the case of the kinematic approximation for the induction equation, the magnetic filaments formation is shown to occur in areas with a hyperbolic velocity profile.
\end{abstract}

\maketitle


\section{Introduction}
 Collapse as a process of the singularity formation for smooth initial conditions represents one of the key issues for understanding  nature for both hydrodynamic turbulence and MHD turbulence.
The Kolmogorov-Obukhov theory \cite{Kolmogorov, Obukhov} of developed hydrodynamic turbulence at
large Reynolds numbers, $ {Re} \gg 1,$ in inertial
interval predicts the divergence of vorticity fluctuations
$\left\langle \delta \omega \right\rangle $ with scale $ \ell $ at
small $ \ell $ like $ \ell^{- 2/3} $, which indicates the connection of Kolmogorov turbulence with
collapse.

Numerous executed in the late 90s
numerical experiments seem to indicate the observation of collapse, with a more accurate examination
showed its absence (a discussion of these issues can be found in \cite{chae2008incompressible, gibbon2008three}).
This problem still remains open, although
there are numerical experiments that show the singularities formation
 on a solid wall in the framework of the three-dimensional Euler equations \cite{hou2007computing}.

  In two-dimensional Euler hydrodynamics, collapse - the appearance of a singularity in a finite time -
is forbidden \cite{wolibner, kato, yudovich}. But this, however, does not exclude the singularity appearance in infinite time
with exponential growth, as evidenced in numerical experiments \cite{KNNR}, in which the formation of the vorticity quasi-shocks is accompanied by exponential in time narrowing their widths. In the three-dimensional Euler hydrodynamics  numerical experiments
also show an exponential increase in time of the  vorticity $\omega $ in the pancake-type vortex structures for which
the exponential in time narrowing of the pancake thickness $\ell$ was also observed
\cite{AgafontsevKuznetsovMailybaev2015, AgafontsevKuznetsovMailybaev2016, AgafontsevKuznetsovMailybaev2017}. The formation of such structures is possible, as
shown in \cite{KuznetsovRuban, KuznetsovRuban2000, Kuznetsov2002}, for  the frozen-in vorticity in the three-dimensional Euler equations and the  vorticity rotor $B$ for two-dimensional  flows \cite {KNNR}. Due to this property, frozen-in vector fields turn out to be compressible. Moreover, it was found that the formation of such structures  can be considered as a folding process when the maximal values of $\omega _{\max }$ and
$B_{\max }$  are evaluated  proportionally to their widths  as
$\ell ^{-2/3}$ \cite{AgafontsevKuznetsovMailybaev2015,AgafontsevKuznetsovMailybaev2016,AgafontsevKuznetsovMailybaev2017},
\cite{KuznetsovSereshchenko2019}. In MHD at high values of magnetic Reynolds numbers, $Re_{m}\gg 1$, the magnetic field can be considered also as a frozen-in field.   Therefore  one can expect,  that the exponential in time growth also should be observed due to compressibility of magnetic field lines. Compressibility of
magnetic field lines as applied to this problem was discussed
for the first time in the works \cite{KuznetsovRuban2000,KPS}.
In particular, \cite{KPS} suggested that frozenness of
magnetic field may cause collapse. In these notes, being  a short review,
we  discuss such possibility for MHD equations in the so-called kinematic
approximation when the magnetic field is relatively small and the inverse
influence of a growing magnetic field on the velocity field can be neglected.
This model is very popular in the theory of turbulent dynamo (see \cite{chertkov1999small,
schekochihin2004simulations,Sokoloff2015} and references there) when
a random velocity field is given. In this work we
consider mainly the case when the velocity field is
regular. Such situation is realized in the convective zone of the Sun.
Observations of convective cells showed that the magnetic field in
this area is heavily filamentated. This was first addressed
attention in Parker's pioneer work \cite{parker1963kinematical}
(see also his book \cite{parker1979cosmical} and the references therein). In particular, in his first work, Parker found the behavior of a magnetic field in
two-dimensional velocity field for convective flow in the case of periodic
cell lattices in the form of rolls.

We discuss this problem statement in
our work. The main summary that can be made is
that for stationary two-dimensional flows magnetic field filamentation
and its exponential growth are due to the presence in hyperbolic flows
areas in the
Okubo-Weiss meaning \cite{okubo1970horizontal, weiss1991dynamics}.
In these areas, the magnetic field is gathered due to frozen-in
magnetic field in a small neighborhood near a stationary hyperbolic
point where velocity vanishes. This process has
exponential in time character and stops due to frozenness
destruction because of finite magnetic viscosity. As a result
the magnetic field is saturated amplifying with respect to the initial field in
$ Re_{m}^{1/2} $ times, that was established in a number of papers (see \cite{weiss1991dynamics,galloway1981convection,stix2002sun}).

The outline of this review is the following. Firstly, we discuss the parameters of the region of convective cells in
the Sun, which occupies the upper part of the convective zone including the lower part of the photosphere. Parameters
of flows in convective cells
allow one to consider the problem of filamentation of magnetic fields in
this area of the Sun in the kinematic approximation. In the third section
we turn to the
magnetic lines representation \cite{KuznetsovRuban2000,KPS},
 analogous to the vortex lines representation, introduced first for
Euler equations in \cite{KuznetsovRuban} (see also
\cite{Kuznetsov2002}). This representation clearly demonstrates
compressibility of magnetic field lines. The velocity of field lines coincides with the fluid velocity component
normal  to the magnetic field direction. Divergence of this velocity component in a general situation is not zero, which ultimately leads to compressibility
continuously distributed magnetic field lines. Based on this fact only it is possible qualitatively to show
 how  filamentation of magnetic field develops in a convective cell. This process happens in the
hyperbolic flow regions, and respectively the focusing of magnetic
field lines takes place  in a small neighborhood of a stationary hyperbolic
points. In the three-dimensional geometry, this process in convective cells
should lead to the formation of filaments flattened relative to
interfaces between convection cells. In the last section we discuss
the question of saturation of the magnetic field due to the finiteness of magnetic viscosity.

\section{Convection in astrophysics}
Numerous observations of the magnetic field spatial  distribution in
solar convective cells (see, for example, data
SOHO~\cite{SOHO} missions, as well as the first observational data
from the most powerful solar telescope DKIST~\cite{DKIST})
indicate a very inhomogeneous distribution of magnetic
fields already within the same convective
cell: the magnetic field is concentrated in the form of magnetic filaments (often
called magnetic flux tubes), the field in which
exceeds significantly the average magnetic field $B_{0}$ on the Sun.
 Especially it is seen
in areas of dark spots~\cite{solanki2003sunspots}.
According to numerous data, $B_{0}$ has a value of several
gauss (see, for example, \cite{bose2018variability} and references therein). (In this paper, we will assume $B_{0} \sim 10 \mbox{ G}$ for estimates.) In each convective cell in its center - in the region of the upward
flow - the magnetic field is practically absent, it is concentrated in
areas of downward flows in the form of filaments with a magnetic field of the order
one kilogauss or more. It should be noted that reconnection of
magnetic flux tubes and various related phenomena in the form of
flares happen higher - in the upper layers of the solar atmosphere,
mainly in the chromosphere and corona (see, for example, the book
\cite{ryutova2015physics} and references there). From this point of view, the question of
the appearance of magnetic filaments seems to us very important and
relevant.

As for convective cells, their horizontal size $L$ according to
observations is about 500-1000 km. In accordance with theoretical and
experimental data about laboratory convection
\cite{gershuni1977convective}  (see also \cite{kuznetsov1980weak})
vertical size of the cells is the same order as their horizontal size, which we
will consider completed
for solar convective cells, that is the most common assumption in these
research. It should also be said about speeds and
densities in area of convective cells. According to
measurements \cite{SOHO}, as well as to many other data (see, for example, \cite{eddy1979new}) the velocity $v$ in a cell is of the order of $1000 \mbox{ m/s} $, and
 density $\rho$ in the photosphere is of the order of $ 10^{-7} \mbox{ g/cm}^{3}$.  Note also that between the convective zone and the photosphere there is no sharp
gradient in density: the density changes smoothly (see, for example,
review \cite{kosovichev2009photospheric} and references therein). The density value
 in the convective zone, of course, is larger  $10^{-7} \mbox{ g/cm}^{3} $. In literature the value of density at the boundary between convective zone and  photosphere
is commonly accepted to have  the order of $10^{-6}-10^{-5} \mbox{g/cm}^{3} $.
Nevertheless, within a
convective
cell the density can be considered practically uniform and,
accordingly, the flow in the cell itself could be considered
incompressible: $\mbox{div}\,\mathbf{v}=0$.

The main issue addressed in this review is a qualitative explanation of the
 observation facts, namely, (i) why the magnetic field filamentation occurs near interfaces of convective cells corresponding to
regions with  downward flows and (ii) why in the central cell regions - areas of upward
flows - the magnetic field is practically missing.
  In this sense, downward convective flows for
magnetic field lines act as peculiar
attractors. The main reason for this phenomenon, as will be shown in
this work, is connected with frozenness of
magnetic field lines into  plasma which is a property  for
zero magnetic viscosity. According to all known data (see,
e.g. \cite{schekochihin2004simulations,Sokoloff2015} and references there) in convective
cells the magnetic Reynolds number $Re_{m}$ is of order $10^{6}$, which
allows ones to neglect in the main approximation by  magnetic
viscosity $\eta_{m}=c^{2}/4\pi\sigma$ term in the MHD equations, where $\sigma$ is
conductivity.

Another important parameter of solar convection is ratio  $\Theta$ between  kinetic
energy density $\rho \mathbf{v}^{2}/2$ and magnetic
energy density $\mathbf{B}^{2}/(8 \pi) $. For example, for the photosphere with
a density of $10^{-7} \mbox{ g/cm}^{3} $ this
ratio turns out to be of the order of $10^{2}$, where for estimates
we took $B_{0}=10 \mbox{ G} $, and the velocity is $ 1000 \mbox{m/s} $. When approaching
the convective zone $\Theta$ becomes about $10^{3}-10^{4}$.
For such ratios energy densities  the magnetic field weakly affects convection and accordingly, the velocity field can be considered the given.  In what follows, we will consider purely stationary flows, moreover, two-dimensional ones, that, in our opinion, is unprincipled for explaining the effect itself.

Further, we restrict our consideration
by purely stationary flows, moreover, two-dimensional flows, such that
in our opinion, is unprincipled to explain the effect itself.

Using only these two assumptions, i.e. high magnetic
Reynolds number and the weak influence of the magnetic field on convection, it will be shown
that the magnetic field in the cell  only due to
convective flow tends to a filamentous state in the form
magnetic flux tubes that are formed in the downward flow region and
parallel to this flow.
 In the region of upward flow, the magnetic field has a tendency
to be vanished; it is pushed
to the periphery of convective cells. The main convection cell model,
 analytically
and numerically investigated in this paper, is the two-dimensional one with flow in
the form of periodic set of rolls  \cite{gershuni1977convective,kuznetsov1980weak}.
In this case, the magnetic field is shown to condensate in the region of downward flow, i.e.,
filamentation takes place
on the interface between cells. Moreover, the magnetic field $B$
due to frozenness only, grows exponentially with time with simultaneous
exponential narrowing of the filament itself. Magnetic field growth and
accordingly, the narrowing of the magnetic filament, as shown in this paper,
stops due to the destruction of the magnetic field frozenness because of
magnetic viscosity. As a result, the magnetic field is saturated in
filament at the level of $B_{0}Re_{m}^{1/2}$.
With $B_{0} = 10\mbox{ G}$ and $Re_{m} = 10^{6} $, the saturation field is of $10^{4} \mbox{ G}$,
which corresponds to observational data. Mechanism that leads to
the formation of magnetic filaments for two-dimensional convective
flows, qualitatively remains almost the same for three-dimensional
convection cells. Magnetic filaments in this case should
flatten while their  growing in the vicinity of the downward flow.
Numerical experiments performed in \cite{Getling13} for
hexagonal cells, as well as recent observations \cite {DKIST} support
the filamentation mechanism discussed here.

It is worth noting  the importance of other magnetohydrodynamic processes arising in astrophysics, and their relationship with convection. Magnetic fields play an important role for accretion disks formed near massive objects, such as black holes, neutron stars, white dwarfs, etc~\cite{Shakura73,Rudiger95,Moss16}. The effect of convection on the magnetic field evolution  has been discussed for a long time~\cite{Hawley01,Narayan02} and it shows the relation between convective motions and the occurrence of magnetorotational instability in accretion disks.  Possibility of suppressing magnetic field reversals in accretion disks due to accretion was discussed in \cite{Coleman17}.
Also we mention works \cite{Ghasemnezhad17, Bethune20} devoted to modeling the relationship between convection and magnetic field in accretion disks. The effect of convective flows is essential for the evolution of magnetic fields in galaxies. Usually, convective flows are directed perpendicular to the equatorial plane. In \cite{Shukurov06, Sur07, Mikhailov13}, magnetic field generation was investigated taking into account the advection of the magnetic field helicity, which is an integral of motion in ideal magnetic hydrodynamics. In \cite{Braithwaite12} there were performed studies of the effect of convection on the magnetic flux while star formation.  All of these  examples indicate the importance of studying convective flows and their effect on the magnetic field from the astrophysical point of view.

\section{Compression of magnetic lines and attractor}

As formulated in the Introduction, in the case of a large density of
kinetic energy versus magnetic energy density
magnetic field is described by the equation of induction in the MHD approximation
for a given velocity field:
\begin{equation}
\frac{\partial \mathbf{B}}{\partial t}=\mbox{rot}\lbrack \mathbf{v}\times
\mathbf{B}]+\eta _{m}\Delta \mathbf{B},\,\,\mbox{div}\,\mathbf{v}=0.
\label{MHD}
\end{equation}
In this case, all the equations are written in dimensionless units:
distances are measured in their typical values of $L,$ velocities - in
characteristic values of $V$. In the case of large magnetic Reynolds numbers
$Re_{m} \gg 1 $ this equation transforms into the frozenness equation:
\begin{equation}
\frac{\partial \mathbf{B}}{\partial t}=\mbox{rot}\lbrack \mathbf{v}\times
\mathbf{B}],\,\,\mbox{div}\,\mathbf{v}=0.  \label{MHD1}
\end{equation}
Due to the vector product in the right-hand side of (\ref{MHD1}) only
velocity component $\mathbf{v_{n}}$, normal to the magnetic field  line,
can change the magnetic field. Tangential component
$\mathbf{v_{\tau}}$ in this case plays a passive role,
providing an
incompressibility condition $\mbox{div}(\mathbf{v_{n}} + \mathbf{v_{\tau}}) = 0$.
Remind that frozenness of magnetic field means that each
Lagrangian particle is pasted to its own field line  and cannot
leave it. A particle, therefore, has only one degree of freedom - motion along a magnetic field that obviously does not change
the field. Hence it immediately follows that $\mathbf{v_{n}}$ is
the velocity of motion for the magnetic field line itself. This fact has
simple geometric explanation. If one considers an arbitrary
curve, then any deformation along the curve, obviously, does not change it; only the deformations normal to
this curve are responsible for its shift.

On the other hand, in the general situation
$\mbox{div}\mathbf{v_{n}} \neq 0$, according to~\cite{KPS}, this is
the reason that continuously distributed magnetic field lines
are compressible objects. In particular, it follows if one considers
 the Lagrangian trajectories specified by the velocity $\mathbf{v_{n}} $
$\mathbf{v_{n}}$, namely by the velocity of magnetic
 field lines   :
\[
\frac{d\mathbf{r}}{dt}=\mathbf{v_{n}}(\mathbf{r},t)\,\,\mbox{}\,\,\mathbf{r}%
|_{t=0}=\mathbf{a}.
\]
The solution of these equations $\mathbf{r}=\mathbf{r}(\mathbf{a},t)$
gives a compressible mapping. The latter follows from the equation for Jacobian $J=\mbox{det}(\partial x_{i}/\partial a_{k})$:
\begin{equation}
\frac{d{J}}{dt}=\mbox{div}\,\mathbf{v_{n}}\,\cdot J.  \label{jacobian}
\end{equation}
It is necessary to emphasize that the equation (\ref{MHD1}) can be integrated in terms of
$\mathbf{r}=\mathbf{r}(\mathbf{a},t)$ (see \cite{KuznetsovRuban, Kuznetsov2002, KPS}):
\begin{equation}
\mathbf{B}(\mathbf{r},t)=\frac{(\mathbf{B}_{0}(\mathbf{a})\nabla
_{a})\mathbf{r}(\mathbf{a},t)}{J}.  \label{Cauchy}
\end{equation}
Here $\mathbf{B}_{0}(\mathbf{a})$ is the distribution of the magnetic field at $t=0$. $\mathbf{B}_{0}(\mathbf{a})$  plays the same role as the Cauchy invariant in ideal hydrodynamics \cite{KPS,yakubovich2001matrix, Kuznetsov2002,
ZakharovKuznetsov1997}.

From the equation (\ref{jacobian}) it follows that the Jacobian $J$ in the general
situation, $\mbox{div}\mathbf{v_{n}}$, can take
arbitrary values, in particular vanishes, when
according to (\ref{Cauchy}) the magnetic field becomes infinite
large. Due to this property, as well as the fact that
the $\mathbf{v_{n}}$ represents
the velocity of the magnetic field lines, it becomes clear that
magnetic field transfer with the speed $\mathbf{v_{n}}$ will be
carried out until the normal component is converted to
zero, i.e. when $\mathbf{B}\Vert \mathbf{v}$.
Areas where $\mathbf{B} \Vert \mathbf{v}$ should represent for
magnetic field peculiar attractors. In three-dimensional
space, as will be discussed below, these attractors must be
two-dimensional. If the velocity field is stationary, then on
attractor, the magnetic field can reach large values, perhaps even
infinite if magnetic viscosity is neglected.

Figure 1 shows schematically the movement of magnetic field lines
(marked in red) in the convection cell velocity field. The speed lines are blue and the arrows
indicate the direction of flow. At the intersection of red and blue lines
black arrows are drawn  showing direction of
the movement of magnetic lines. These arrows are perpendicular to the magnetic
lines, show the direction of the normal velocity component
 $\mathbf{v_{n}} $.  It immediately follows from this simple drawing that in the strip $ 0 \geq y \geq - \pi / 2 $ all the magnetic lines from the right cell move closer to the line $ x = 0 $ on the left, and all
magnetic lines from the left cell - to this line to the right. Thus, convective flow gathers all the  magnetic field lines on the line $ x = 0 $. We emphasize that such a process is possible due to the compressibility of continuously distributed
magnetic lines. Convective flow rakes all the magnetic lines, forming a magnetic filament. In this case, as is easily seen from this figure, the largest magnetic field should arise in the vicinity of the point $ x = y = 0 $. This  point for convective flow is hyperbolic. Just because of hyperbolicity, as it will be
shown in the next section, filament formation becomes possible.
Note that raking the magnetic field occurs due to the
magnetic field frozenness.

\begin{figure}[ht]
\centering
\includegraphics[width=0.49\textwidth]{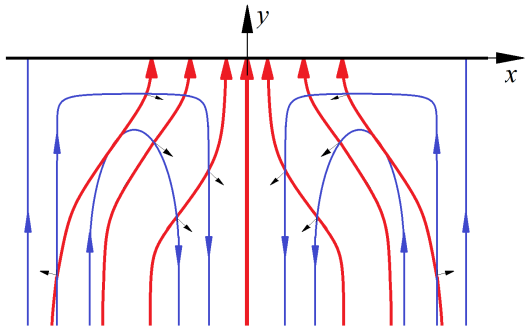}
\caption{Magnetic field lines in convective cell.}
\end{figure}

\section{Convective cell and boundary conditions}

Let us discuss how the velocity field is arranged in one convective cell. In the center of the cell there is an upward flow, and at its border -
moving stream down. In the simplest two-dimensional geometry, the velocity in
the cell can be represented through the stream function $\psi$: $v_{x} = -\partial_{y} \psi, \, \, v_{y} = \partial_{x} \psi $. For periodic
chain of rolls with circulation, changing its sign when moving from one
cells to another, $\psi $ can be written as a product of
two sines: $\psi
=C\sin \left( k_{1}x\right) \sin \left( k_{2}y\right) $. So,
\begin{eqnarray*}
v_{x} &=&-Ck_{2}\sin \left( k_{1}x\right) \cos \left( k_{2}y\right) \\
v_{y} &=&Ck_{1}\cos \left( k_{1}x\right) \sin \left( k_{2}y\right)
\end{eqnarray*}

For a stationary system for Benard convection $k_{1} \neq k_{2} $, but
they are of the same order~\cite {gershuni1977convective}
(see also \cite{kuznetsov1980weak}). As it will be seen below, for everything
subsequent it is unprincipled. Further therefore for simplicity
put $k_{1} = k_{2} = 1$, and the constant $C = 1$ (this corresponds to the transition
to dimensionless variables). As a result
\begin{equation}
\psi =\sin x\cdot \sin y,  \label{psi}
\end{equation}
and velocity components
\begin{eqnarray*}
v_{x} &=&-\sin x\cdot \cos y,  \label{x} \\
v_{y} &=&\cos x\cdot \sin y.  \label{y}
\end{eqnarray*}
Moreover, the line $y=0$ will be considered the upper boundary
convection cells. The velocity along the boundary in this case is parallel
surface, the normal component is respectively zero.

In  numerical experiments, we present the results for two cells with one
 common interface $x=0$ along which the cooled fluid moves
down. Two cells correspond to a rectangular
area [$ - \pi \leq x \leq \pi, \, 0 \geq y \geq
- \pi $]. Along the lines $ x = \pm \pi, \, 0 \geq y \geq - \pi $ fluid
pops up (upflows), and along the center line
$x = 0, \, 0 \geq y \geq - \pi $ goes down (downflow). If in
the initial moment of time the magnetic field $ \mathbf{B_ {0}} $ is directed along
$y$-axis , then one can see that the normal velocity component
 relative to $\mathbf{B_ {0}} $ will have positive
$x$ -projections in the upper right corner in the left cell, i.e. to
lines $x = 0, \, 0 \geq y \geq \pi / 2 $ and, respectively, negative
$ x $ projections in the left corner of the right cell. Thus the magnetic
field lines  in this area will be drawn to the downward
flow, i.e. to the center line where the filament should be formed.
A similar picture should occur in the lower left and, accordingly,
the right corners of the rectangle.

\section{Filamentation}

We turn now to finding the magnetic field depending on
time and coordinates. This problem is in the general formulation for
equation (\ref{MHD1}) is as follows. In two-dimensional
geometry when the magnetic field lies in the plane of convective
flow ({x}, {y}), we introduce the magnetic potential,
presenting it in the form
\begin{equation}
A=-B_{0}x+a,  \label{fluct}
\end{equation}
where the magnetic field $ \mathbf{B_{0}} $ is assumed to be uniform,
directed along the $y$ axis, and the fluctuation of $a$ is periodic
coordinate function both in $x$ and in $y$.
 The latter  automatically
will retain the full flux of the magnetic field across the boundary $ y = 0 $.
For initial time, we can assume no fluctuations, $a(t=0)=0$,
i.e.  we start with a uniform magnetic field.

As explained above (see Fig. 1), magnetic field filamentation should
appear in the upper right corner of the left cell near the top of the down
stream at $x = y = 0$. The growth of the magnetic field at the maximum point should be
observed precisely near this point.

The equation for the magnetic potential $ A $ is obtained by integrating equations for frozen-in field (\ref{MHD1}):
\begin{equation}  \label{magnetic}
\frac{\partial A}{\partial t}+(\mathbf{v}\cdot \nabla) A=0,
\end{equation}
where the magnetic field can be  expressed as follows
\begin{equation}  \label{field}
B_x= \frac{\partial A} {\partial y}, \,\, B_y=-\frac{\partial A} {\partial
x}.
\end{equation}
From these relations it follows that the lines of constant value of $ A $ coincide
with a magnetic field line. From the equation (\ref{magnetic}), with
on the other hand, by the scalar product $ (\mathbf{v} \cdot \nabla
A) $ it can be seen that (\ref{magnetic}) includes only the normal
velocity component to the equipotential $ A = \mbox{const} $, i.e. to magnetic
field line.

The equation for $ A $ is simply integrated
using the method of characteristics. The equation for the characteristic has the form
\[
\frac{d\mathbf{r}}{dt}=\mathbf{v}(\mathbf{r})
\]
with the initial conditions $ \mathbf{r}|_{t = 0} = \mathbf{a} $.
This equation in the component version represents the Hamilton equations
\begin{equation}
\frac{dx}{dt}=-\frac{\partial \psi }{\partial
y},\,\,\frac{dy}{dt}=\frac{\partial \psi }{\partial x}  \label{Hamilton}
\end{equation}
with the initial conditions $ x (t = 0) = a_{x} $ and $ y (t = 0) = a_{y} $.
The coordinates $ x $ and $ y $ in these equations are canonically
conjugate quantities, and the stream function $ \psi (x, y) $ is
Hamiltonian. Since the velocity field is independent of
time, then $ \psi (x, y) $ is the conserved quantity.
Wherein the
magnetic potential on the characteristic
does not change with time. Thus, the dynamics of the system (\ref{Hamilton})
is determined by the properties of the Hamilton function $ \psi (x, y) $. For a finite area
the function $ \psi (x, y) $, as a two-dimensional relief, is characterized by its extremums -
minima, maxima and saddle points. At extremum points
the gradient from $ \psi(x, y) $ is equal to zero, which corresponds to zero velocity.
Which kind of extreme point
is determined from the expansion of $ \psi (x, y) $ in the vicinity of the extremum $ {\bf r} = {\bf r_0} $:
\begin{equation}
\psi({\bf r})= \psi({\bf r_0}) +\frac 12 D_{ij}\Delta x_i \Delta x_j + ...,\label{extremum}
\end{equation}
where $\Delta {\bf r}= {\bf r}-{\bf r_0}$,
\[
D_{ij}=\frac{\partial^2\psi }{\partial x_i \partial x_j}|_{\bf r=r_0}.
\]
At  maximum or minimum, the quadratic form $ D_{ij} \Delta x_i \Delta x_j $ is sign-definite. At this point
the eigenvalues of the matrix $ D_{ij} $ are sign-definite if the following inequality is fulfilled
\begin{equation}
\label{Okubo}
\psi_{xx}\psi_{yy}-\psi^2_{xy}>0.
\end{equation}
According to \cite{weiss1991dynamics, okubo1970horizontal} such points are called elliptic, respectively,  the region where
the inequality (\ref{Okubo}) takes place is  elliptic. With a opposite sign in the inequality (\ref{Okubo}), the stationary point becomes hyperbolic, and the corresponding region with the opposite  sign in (\ref{Okubo}) is called hyperbolic.

Along the characteristic $x(t), \, y(t) $, the potential $A$ is constant, representing a passive scalar. Depending on the region (elliptic or hyperbolic), the potential $A$ will behave differently.

So, for example, for the stream function (\ref{psi}), the point $ x = y = 0 $ is obviously hyperbolic; to this point, according to Fig. 1, the magnetic field should be attracted, which leads to filamentation of the magnetic field. The point $ x = y = - \pi / 2 $, on the contrary, is an elliptical point around which the vector potential will rotate.

Now consider the solution to the problem for the initial condition (\ref{fluct}) for the stream function (\ref{psi})
Due to the fact that the magnetic
 potential fluctuations $ a $ at $ t = 0 $ are absent, the magnetic potential on the characteristic
will be equal to $ A = -B_{0} a_{x} $, i.e. depends only on the initial
 values of the $ x $ -coordinate of the fluid particles $ a_{x} $.

Equations (\ref{Hamilton}) are simply integrated. From
the equality $ \psi (x, y) = \psi (a_x, a_y) $  one can find
dependence $ y = y (x, a_x, a_y) $ and then
substitute it in the right-hand side of the first
equations (\ref{Hamilton}). As a result, the equation for $ x $
\[
\frac{d x}{dt}=v_{x}(x,a_{x},a_{y})
\]
is integrated trivially. Thus, we come to the general  solution
of the Cauchy problem for equation (\ref{magnetic}); this solution is written
implicitly.

We will be interested in the behavior of the maximal magnetic field.
The qualitative considerations given above show that the maximal
magnetic field should be in a small neighborhood of
point $ x = y = 0 $, i.e. for the center of the beginning of the downward flow, at the border
between cells. We can take small deviations from the point $ x = 0 $, $ y = 0 $,
considering $ x $ and $ y $ small. For such values of $ {x} $ and $ {y} $, the stream function,
according to (\ref{psi}), can be approximately written as
\[
\psi ={x}{y}.
\]
The initial condition for $\psi $ is the following
\[
\psi ={a_{x}}{a_{y}}.
\]
For such a stream function, the equation
for $ {x} $ transforms into the linear one:
\[
\frac{dx}{dt}=-{x},
\]
whose solution gives an exponential
narrowing of the scale
\begin{equation}
x=a_{x}e^{-t}.  \label{compression}
\end{equation}
For $ {y} $ we have exponential
growth: $ y = a_{y} e ^ {t} $. From these asymptotics one can find the
magnetic field behavior in this area. To do this, express
derivatives
with respect to $ x $ and  $ y $ in the equations (\ref{field}) through derivatives with respect to
variables $ a_{x}, a_{y} $. The easiest way to do this is
based on the Jacobian technique. Note that due to the Hamiltonian
equations (\ref{Hamilton}) Jacobian
\[
\frac{\partial (x,y)}{\partial (a_{x},a_{y})}=1.
\]
In particular,
for $B_{x}$ we have the following chain of transformations:
$$B_{x}=\frac{\partial (A,x)}{\partial (y,x)}=$$
\begin{equation}
\frac{\partial (A,x)}{\partial
(a_{y},a_{x})}=\frac{\partial A}{\partial a_{y}}\frac{\partial x}{\partial
a_{x}}-\frac{\partial A}{\partial a_{x}}\frac{\partial x}{\partial a_{y}}.
\label{B-x}
\end{equation}
Similar
calculations for $ B_{y} $ give
$$B_{y}=-\frac{\partial (A,y)}{\partial (x,y)}=-\frac{\partial (A,y)}{\partial
(a_{x},a_{y})}=$$
\begin{equation}
-\frac{\partial A}{\partial a_{x}}\frac{\partial y}{\partial
a_{y}}+\frac{\partial A}{\partial a_{y}}\frac{\partial y}{\partial a_{x}}.
\label{B-y}
\end{equation}

Hence for asymptotically as
${x\rightarrow 0}$ and ${y\rightarrow 0}$ we find
\begin{equation}
B_{x}=0,\,\,B_{y}=B_{0}e^{t}.  \label{filam}
\end{equation}
If at the initial moment of time the $ x $-component of the magnetic
field is not equal to zero, then $ B_{x} $ decays exponentially with time (see
below). Thus, the maximum value of the magnetic field increases during
time exponentially in the neighborhood of
hyperbolic points. It is important that the maximal field is directed along
downstream.

\section{Numerical modelling}

In the numerical integration of the MHD equations, at first
magnetic potential fluctuations (\ref{fluct}) for $ a $ were determined
from  equations (\ref{magnetic}):
\begin{equation}
\frac{\partial a}{\partial t}+(\mathbf{v}\cdot \mathbf{\nabla })a=v_{x}B_{0}.
\label{aeq}
\end{equation}
This problem was solved in the region $ - \pi <x <\pi, $ $ - \pi <y <$ $ \pi $ with
periodic boundary conditions along both coordinates, taking
further only the lower strip ($ -\pi <y <0 $)  corresponding to convective
cells. For known magnetic potential  we calculated
a magnetic field by means of formulas (\ref {field}). For convenience, the average magnetic field
$ B_{0} = 1$ was chosen.

Numerically, the equation (\ref{aeq}) was solved on a grid  $ 2000 \mbox {x} 2000 $ using an explicit numerical scheme with a small step, providing
algorithm stability~\cite{SamarskiiPopov1992}. The time step was selected in accordance
with a spatial step; in most cases, $ \Delta t = 2.5 \cdot 10^{-5}$.

Figures 2--4 show the results of integration for the vector
potential $ A (x, y) $ at three moments in time. Recall that
level lines $ A (x, y) $ coincide with the  magnetic field lines,
which in the figures correspond to the boundaries between the regions of one
color. One can see the twist of the magnetic field lines
 inside the cells over time, which indicates that the corresponding  region is elliptical.
Moreover, in the neighborhood of the $ y $-axis  isolines of $ A (x, y) $ become with time more dense
 that corresponds to
an increase of the magnetic field in the center, i.e. in the region of downward flow.

\begin{figure}[ht]
\centering
\includegraphics[width=0.49\textwidth]{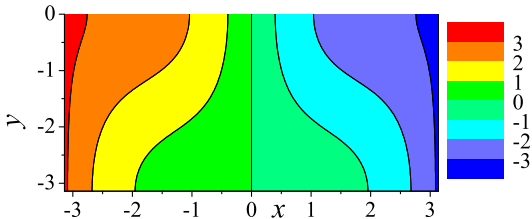}
\caption{Vector potential of the magnetic field in ideal case for $t=1$.}
\end{figure}

\begin{figure}[ht]
\centering
\includegraphics[width=0.45\textwidth]{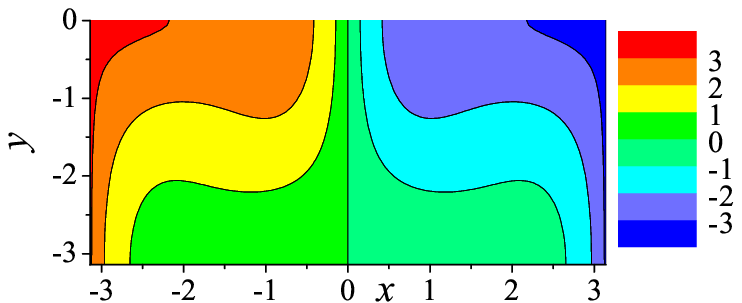}
\caption{Vector potential of the magnetic field in ideal case for $t=2$.}
\end{figure}

\begin{figure}[ht]
\centering
\includegraphics[width=0.49\textwidth]{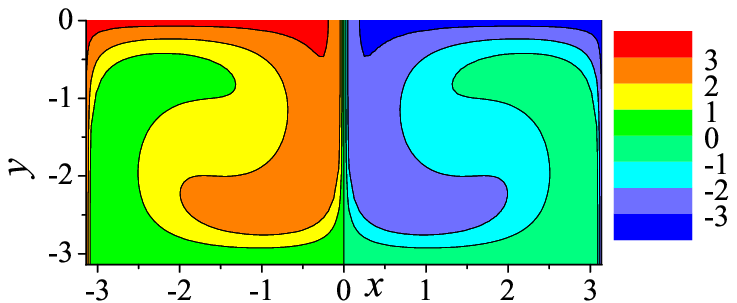}
\caption{Vector potential of the magnetic field in ideal case for $t=5$.}
\end{figure}

\begin{figure}[ht]
\centering
\includegraphics[width=0.49\textwidth]{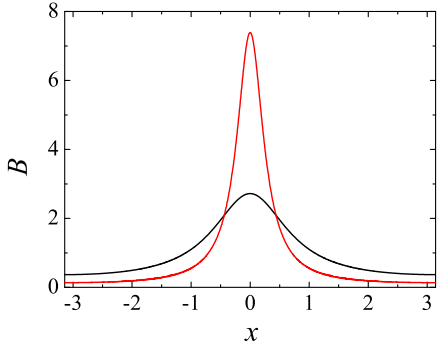}
\caption{Magnetic field on $x$ axis. Black line shows case $t=1,$ red -- $t=2$.}
\end{figure}

\begin{figure}[ht]
\centering
\includegraphics[width=0.49\textwidth]{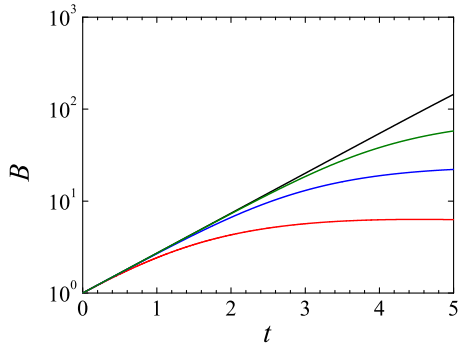}
\caption{Maximal magnetic field evolution. Black line shows ideal case, red -- $Re_{m}=10,$ blue -- $Re_{m}=10^{2},$
green -- $Re_{m}=10^{3}$.}
\end{figure}

Fig. 5 shows the coordinate dependence of the vertical
magnetic field components on the ${x}$ axis. Note that the vector potential level
lines there are directed vertically when the horizontal
 component of the magnetic field vanishes. On the ${x}$ axis
the magnetic field peak is formed at the point ${x=0}$, which width
decreases as  the peak grows.

The maximal magnetic field in the filament increases in time
exponentially with the growth rate  $\gamma=1$ (Fig. 6 and
7, black curves), in full agreement
with  analytic expression (\ref{filam}).

\section{Influence of magnetic viscosity on the magnetic field evolution}

The growth of the magnetic field in the filament is associated with its frozenness, which is
destroyed due to finite magnetic viscosity.
In this case, in the equation for $ a $
one needs to add the term responsible for magnetic viscosity:
\[
\frac{\partial a}{\partial t}+(\mathbf{v}\cdot \mathbf{\nabla
})a=B_{0}v_{x}+\frac{1}{Re_{m}}\Delta a, \]
where $Re_{m}$ --
is the magnetic Reynolds number (in this equation we use dimensionless variables).

The corresponding spatial distribution of the vector potential
for the case $Re_{m}=10^{2}$  is shown in Fig.~8. No cardinal
changes occur compared with the non-dissipative case,
except a little less twist.

However, studying evolution of the
magnetic field with time for different magnetic Reynolds numbers
 and its comparison with the case of zero magnetic viscosity is much more interesting.
At the early stage, when frozenness works, the magnetic field increases exponentially, and then, with a decrease in the filament thickness, saturation occurs reaching the stationary value $ B_ {sat} $ due to the frozenness destruction.  Moreover, with the magnetic Reynolds number decrease this value becomes smaller (Fig. ~ 6).

To estimate $B_{sat}$, we return to the original equation for the magnetic
field (\ref{MHD}), written in the dimensionless form:
\begin{equation}
\frac{\partial \mathbf{B}}{\partial
t}=\mathrm{rot}[\mathbf{v},\mathbf{B}]+\frac{1}{Re_{m}}\Delta \mathbf{B}
\label{MHDdimless}
\end{equation}
Both the numerical experiment and analytical calculations show, that on the line $ y = 0 $ the magnetic field is maximal at the point $ x = 0 $. In the vicinity of this point, a magnetic field (having only one
vertical component) obeys the equation that follows from
(\ref{MHDdimless}):
\[
\frac{\partial B_{y}}{\partial t}=\frac{\partial }{\partial x}\left(
xB_{y}\right) +\frac{1}{Re_{m}}\frac{\partial ^{2}B_{y}}{\partial x^{2}}.
\]
Stationary solution to this equation is found from integration
\[
\frac{\partial }{\partial x}\left( B_{y}x\right)
+\frac{1}{Re_{m}}\frac{\partial ^{2}B_{y}}{\partial x^{2}}=0,
\]
that gives
\begin{equation}
B(x)=B_{max}\exp \left( -\frac{(x-x_{0})^{2}}{2Re_{m}}\right) .  \label{Bst}
\end{equation}
Obviously, the integration constant $x_{0}$ should be
set to zero, and the value of $B_{max}$ should be determined from the conservation of
magnetic field flux. From (\ref{Bst}) we have
\begin{equation}
\Phi \approx B_{max}\sqrt{\frac{2\pi }{Re_{m}}}.  \label{integr1}
\end{equation}
At the initial moment, $\Phi=2\pi B_{0}$, whence the estimate for
maximal amplitude is of
\[
B_{max}=B_{0}\sqrt{2\pi Re_{m}}.
\]
It is worth noting that a similar estimate was obtained in a number of works~\cite{weiss1991dynamics,galloway1981convection,stix2002sun}. Unfortunately, it is rather difficult to determine when such estimate was first received.
Note that the estimates for the magnetic field are close to the results of
numerical simulation (Fig.~6, 9). The width $\delta$ of this distribution
determined only by magnetic viscosity. In dimensional
variables $\delta \sim L / \sqrt{Re_{m}}$. Saturation time of passing to stationary value can be evaluated as $T\sim \frac{1}{2}\ln Re_{m}$. In
dimensional variables saturation time
\[
t_{0}=T\frac{L^{2}}{\eta _{m}}=\frac{1}{2}\frac{L^{2}}{\eta _{m}}\ln \left(
\frac{LV}{\eta _{m}}\right) .
\]
\begin{figure}[ht]
\centering
\includegraphics[width=0.45\textwidth]{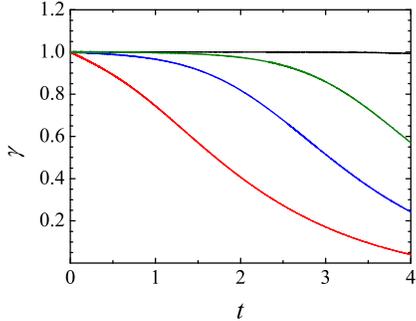}
\caption{Growth rate $\protect\gamma=d(\mbox{ln}B_{max})/dt$. Black line shows ideal case, red -- $Re_{m}=10,$ blue -- $Re_{m}=10^{2},$
green -- $Re_{m}=10^{3}$.}
\end{figure}

\begin{figure}[ht]
\centering
\includegraphics[width=0.49\textwidth]{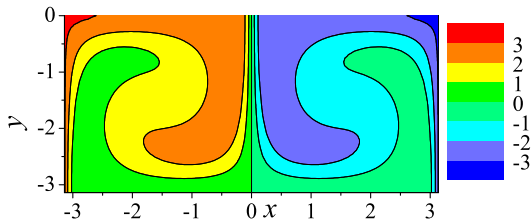}
\caption{Vector potential distribution  for $t=5,$
$Re_{m}=10^{2}$.}
\end{figure}

\begin{figure}[ht]
\centering
\includegraphics[width=0.49\textwidth]{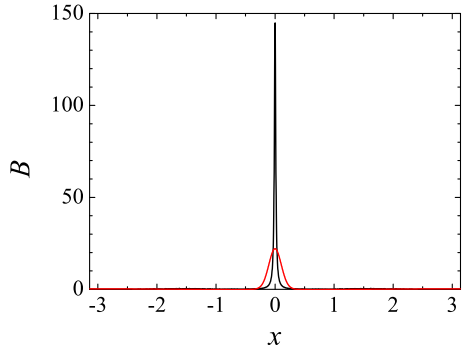}
\caption{Magnetic field on $x$ axis for $t=5$. Black line shows ideal case, red -- $Re_{m}=10^{2}$.}
\end{figure}

\section{Concluding remarks}

Thus, for two-dimensional flows, filamentation of
magnetic field occurs in the downward flow region. An exponential increase of the magnetic field is observed in
neighborhood of the flow hyperbolicity, which in our opinion is
the main criterion of the magnetic field filamentation.
In the case of convective rolls, considered in this paper,
this occurs in the vicinity of the hyperbolic point $x=y=0$, where
velocity components
$v_{x} =-x, \, v_{y}=y$ respectively. Perpendicular
the $y$ axis the velocity component $v_{x}$ due to frozenness gathers
magnetic field, leading to its exponential growth. The question arises,
how sensitive is the influence of the third component of
magnetic field, namely, for a real three-dimensional problem. As noted
magnetic field growth in a two-dimensional situation is determined by hyperbolicity
flows near the interface of two convective cells. In the three-dimensional case, this hyperbolicity persists: when approaching the interface the flow can be considered flat, i.e. it has the very structure
as in the two-dimensional case:
\begin{equation}  \label{2Dvel}
v_{x}=-x,\,v_{y}=y, \, v_{z}=0.
\end{equation}
In this case, according to (\ref{MHD1}) the behavior of the third component
of magnetic field parallel to the interface in its vicinity will be
determined from the equation
\begin{equation}
\frac{\partial B_{z}}{\partial t}+(\mathbf{v}\nabla )B_{z}=0.  \label{MHD2}
\end{equation}
Thus, in a neighborhood of the interface plane, $ B_{z} $ represents
the Lagrangian invariant unchanged when moving with
plasma.

When magnetic viscous dissipation is taken into account, $ B_{z} $ will attenuate. Concerning
 two other components of the magnetic field,  at $\eta=0$ they are found
from the equations
\begin{eqnarray*}
\frac{\partial B_{x}}{\partial t}+(\mathbf{v}\nabla )B_{x} &=&-B_{x}, \\
\frac{\partial B_{y}}{\partial t}+(\mathbf{v}\nabla )B_{y} &=&B_{y},
\end{eqnarray*}
where the velocity is given by (\ref{2Dvel}). From the second equation it follows that
component $ B_{y} $ grows exponentially
(compare with (\ref{filam})), and $B_{x}$, on the contrary, falls
exponentially $ \sim e^{-t} $. This process continues until frozenness is destroyed and the magnetic field in the filament comes out
saturation due to magnetic viscosity. It follows from this fact  that
magnetic field filaments should be flattened relative to the plane
interface. It is also important to note that the difference between vertical and
horizontal components of magnetic field occurs large enough.

We now discuss how the variability of convective cells affects the mechanism of formation of magnetic filaments as well as the magnetic field of  filaments influences on the convective flow.

Characteristic  variability time of convective cells can be estimated as
the ratio of the cell size $L \sim 10^{8} \mbox{ cm}$ to the characteristic
speed $V \sim 10^{5} \mbox{ cm/s}$, which is consistent with the observational
data. This is the time of the order of the inverse growth rate $\gamma^{-1}$, i.e.
the magnetic field amplification in the "quiet" mode is roughly of the
same order as $L/V$, but the convective motion does not stop and therefore
pushing out
magnetic field from the center of the cell to the periphery does not stop.
The second question is to take into account the inverse effect of the magnetic field on the convective flow
characteristics.
For a typical speed of the order of $10^{5} \mbox{ cm/s}$ and
characteristic density $\sim 10^{-5} \mbox{ g/cm}^{3}$ it is possible
to get that up to a magnetic field of the order of 1 kG
it makes sense to speak about a  weak influence of magnetic fields on
flow characteristics. For larger fields, in our opinion,
this mechanism should nevertheless work, since in the center of the cell due to the expulsion of the magnetic field to the periphery, the flow is practically independent of the magnetic field, but in the downward flow region, where
magnetic filament is formed, magnetic field being parallel
to the flow should be practically  stationary.

We also note that the results presented in this paper are generally confirmed by numerical
simulations \cite{getling2001convective, kitiashvili2010mechanism,
Getling13} and observational data ~\cite{Tiwari13, Zakharov08}, indicating a correlation of the amplifying magnetic fields with
downward flows. In addition, it is worth mentioning a number of numerical results ~\cite{Getling06, Getling13} obtained for convective hexagonal cells in which the indicated correlation was observed.

The authors are grateful to V.V. Krasnoselskykh, I.N. Kitiashvili, A.G.
Kosovichev and I.V. Kolokolov for useful discussions and a number of valuable comments.
The work of E.K. was supported by the Russian Science Foundation (grant 19-72-30028). The work of E.M. was performed
within the project "Extreme phenomena and coherent structures in nonlinear physics" of Ministry of High Education and Sciene of RF.

\end{document}